Where Assessment Validation and Responsible AI Meet

Jill Burstein and Geoffrey T. LaFlair

Duolingo, Inc.

5900 Penn Avenue

Pittsburgh, PA 15206

{jill, geoff}@duolingo.com

**Abstract**

Validity, reliability, and fairness are core ethical principles embedded in classical argument-based assessment validation theory (Kane, 2013;1992; Chapelle et al, 2008). These principles are also central to the Standards for Educational & Psychological Testing (AERA & APA & NCME, 2014; the *Standards*) which recommended best practices for early applications of artificial intelligence (AI) in high-stakes assessments for automated scoring of written and spoken responses. Responsible AI (RAI) principles and practices set forth by the AI ethics community are critical to ensure the ethical use of AI across various industry domains. Advances in generative AI have led to new policies as well as guidance about the implementation of  RAI principles for assessments using AI (ATP, 2024; US Department of Education, 2024). Building on Chapelle et al.'s (2008) foundational validity argument work to address the application of assessment validation theory for technology-based assessment, we propose a unified assessment framework that considers classical test validation theory and *assessment-specific* and *domain-agnostic* RAI principles and practice. The framework addresses responsible AI use for assessment that supports validity arguments, alignment with AI ethics to maintain human values and oversight (von Davier A. & Burstein, in press), and broader social responsibility associated with AI use.

**Introduction**

Validity, reliability, and fairness are integral to classical argument-based assessment validation theory (Kane, 2013; 1992; Chapelle et al., 2008) and professional assessment standards (e.g., AERA, APA, NCME, 2014; the *Standards*). The *Standards* offers best practices for the early use of artificial intelligence (AI) in high-stakes assessments, such as automated scoring of written and spoken responses. Responsible AI (RAI) consists of principles and practices to ensure ethical AI use. RAI is shaped by, and grounded in AI ethics  - i.e., the human values associated with AI use across domains (such as fairness and accountability) (von Davier, A. & Burstein, in press).  Developments in generative AI have led to new assessment policy that provides guidance on applying RAI principles in assessments (ATP, 2024; US Department of Education, 2024).

Both the new technological and policy developments are catalysts for revisiting previous work on validity theory as it applies to technology-based assessments. Recently, von Davier, A. & Burstein (in press) explored connections among classical validity argument, human-in-the-loop AI, and responsible AI standards for developing digital assessments. They focussed, more specifically, on how these drivers leverage human–centered AI principles.This paper compliments von Davier A. & Burstein,  focusing on the critical importance of broadening RAI considerations for developing digital assessments.

In the spirit of this volume, this paper honors and expands on Chapelle et al.'s (2008) foundational work regarding assessment validation theory in technology-based assessments.   It emphasizes the need to build a unified assessment framework that considers classical assessment validation and a broader set of RAI principles and practices that are both *assessment-specific* and *domain agnostic.* It further stresses the importance of aligning with AI ethics to maintain human values and oversight  (The World Economic Forum, 2024; von Davier, A. & Burstein, in press), and broader social responsibility associated with AI use..

**Classical Assessment Validation Theory and Responsible AI**

*Inference-based validity frameworks*

A test score represents a snapshot of a person's ability with regard to a target construct (such as listening on a language proficiency test). This score is usually interpreted as an indicator of what a person knows or can do in a target domain (such as listening comprehension in their second language). Further the test score is often used to make a decision, such as admissions to university for international students who want to study in English-speaking countries, in which the decision is based on English language proficiency assessments. The quality of the interpretation and use of these test scores depends on the evidence generated though the test development process that supports their validity, as well as research on the test. There is a general consensus that validity is a property of the interpretations and uses of test scores. In addition, the extent to which test scores are valid depends on the amount and quality of evidence for proposed interpretations and uses of test scores (Chapelle, 2021; Kane, 2013; Messick, 1995). Argument-based validation approaches serve to organize evidence around inferences or assumptions that are made when interpreting and using test scores (Kane, 1990; 2013).

The American Educational Research Association (AERA), American Psychological Association (APA), and National Council on Measurement in Education (NCME) (2014) recommend the following types of evidence be collected to support test validity claims.

- Test content (e.g., analysis of the construct relevance of task types and item topics)

- Cognitive and response processes (e.g., eye-tracking to examine test-taker reading patterns)

- Internal test structure (e.g., DIF analysis to confirm that test items function comparably across test-taker subgroup populations)

- Relationships with external criteria (e.g., relationship between test score and grade point average)

- Test consequences (e.g., examining test score use)

These types of evidence can be used to support various inferences in a validity argument. The number and type of inferences can vary depending on the test purpose, with high-stakes tests requiring more and higher quality evidence. Kane (2013) argued that test score users make at minimum three inferences in standardized assessments: a *scoring* inference, a *generalization* inference, and an *extrapolation* inference. In other words, there are three core assumptions that are made when using a test score. These assumptions begin with the test performance and end with decisions (or some might argue consequences). The *first* assumption is that the scoring materials and processes are appropriate and unbiased. The *second* assumption is that a test score is consistent. This means that if a person takes the same test, or different forms, multiple times (while the trait remains stable) it is expected that their score will remain relatively the same. The *third* assumption is that the test score can be used as a springboard to move beyond interpretations about how a person performs on a test to how they would perform in the target domain. From test performance to score use, these assumptions (and their evidence) build on each other. If an early assumption does not have evidence to support it, the validity argument falls apart. For example, if an automated scoring system has bias for one or more groups or was not trained in a manner such that the scores capture performance on the construct accurately, the validity argument deteriorates regardless of the quality of evidence for other inferences. From the perspective of the scoring inference, such cases illustrate the need for careful responsible creation of rubrics, training of raters, and design of models in the development of automated scoring systems.

 As stated earlier, the number of inferences made when using test scores can vary. Chapelle et al. (2008), when applying Kane's argument-based validation framework to the Test of English as a Foreign Language (TOEFL), added three additional inferences to the validity argument. They determined that the interpretation and uses of TOEFL scores contains a total of six inferences (additions are in bold): *domain definition*, evaluation (scoring), generalization, *explanation*, extrapolation, *utilization*. In expanding on Kane's (2013) three inferences, they argue that the first assumption made when using a test score is that performances reveal abilities that have connections to the same abilities in the target domain (domain definition). In making the explanation inference, it is assumed that performance can be attributed to the relevant construct. Finally, in using test scores, it is assumed that they are meaningful and have positive washback (utilization) (Chapelle et al., 2008). Regardless of the composition of inferences in validity arguments, they should be plausible and supported with sufficient evidence given the proposed interpretation and uses of test scores. As new tools, such as generative AI, are developed and used in test design, it is important to consider how their use contributes to validity arguments.

*Responsible AI*

*AI*[1] systems (henceforth, AI) are designed to generate outputs, such as predictions, recommendations, or decisions for a particular task (The National Institute of Standards and Technology AI Risk Framework, 2023; NIST AI RMF). AI for assessment has been largely used to generate predictions (scores) for automated writing evaluation (AWE) (Shermis & Wilson, 2024; Shermis & Burstein, 2013, 2003) and automated speech scoring (Zechner & Hsieh, 2024; Evanini & Zechner, 2019). However, as a result of the increased availability of large language models (LLMs), automated content generation systems are now used for test development (Attali et al., 2022; Runge et al., in press). The field is exploring the application of LLMs to automated scoring (Naismith et al., 2023; Yancey et al., 2023) and test security, such as to detect malicious behaviors (Niu et al., 2024) and monitor proctoring decisions (Belzak et al., 2024). Such innovations increase AI's footprint in the test development, administration, and scoring process, further increasing the need for guidelines and standards outlining their responsible use.

*Responsible AI* (RAI) is a domain-agnostic field, independent from assessment, that is intended to mitigate potential harms related to AI use. RAI practices are essential to uphold validity for AI-powered assessments, and can be adapted and applied to different domains as designated by domain experts.

This section discusses RAI for assessment and demonstrates how domain-agnostic RAI frameworks can be leveraged to integrate classical validation theory and RAI principles and practice. Doing so ensures that RAI practices are aligned with AI ethics and grounded in prevailing best practices developed by a cross-disciplinary set of experts, including domain experts as well as those from computer science, ethics, and law. Additionally, using a domain-agnostic approach facilitates multi-stakeholder collaboration, for example, discussions between assessment stakeholders and government agencies when developing regulations.

---

[1] Note that in this paper, AI refers broadly to generative AI, supervised and unsupervised machine learning (ML), and statistical modeling techniques.

While AI offers advantages and opportunities for assessment developers and researchers, it also poses risks. RAI provides guidelines for developing and managing AI to mitigate harm and contribute positively to society. Human experts create RAI frameworks to guide the development and application of responsible practices which are often accomplished through human oversight. For assessment, and especially for high-stakes assessment, human oversight for AI use is essential to address problems associated with AI use such as hallucination in automatically generated content (Ji et al., 2023) and potential bias in AI systems (Mihalcea et al, 2024; Belzak, 2023; and, Johnson et al., 2022). These types of risks and associated RAI practices are discussed extensively in the RAI literature.

For AI-powered assessments, integrating RAI practices is essential across the assessment ecosystem for test design, measurement, and security (Burstein et al, 2022). This integration better ensures test validity, since these practices are distributed across validity inferences. For example, as part of test design, and in line with the *explanation* inference, RAI human review practices ensure that AI-generated items and content are construct-relevant. Similarly, in the context of measurement and aligned with the *evaluation* inference, RAI evaluation practices ensure accurate AI-generated score predictions. As part of security, and aligned with the *utilization* inference, RAI practices leverage human proctoring to verify AI decisions about cheating behaviors. The practices of human review, accurate measurement, and test security are not novel. However, we need to make sure that these practices are appropriately modified to handle the unique challenges, especially associated with generative AI (such as hallucination). To that end, recent RAI policy for assessment has recommended guidelines and best practices (see Association for Test Publishers (ATP), 2024; U.S. Department of Education, Office of Educational Technology, 2023, 2024; the International Privacy Subcommittee of the ATP Security Committee, 2021; International Test Commission & Association of Test Publishers (ITC-ATP), 2022; Molina et al., 2024; and OECD, 2023).

NIST AI RMF (2023) is a well-recognized, domain-agnostic RAI framework[2] that has been leveraged for RAI assessment policy (see ATP, 2024; US Department of Educational Technology, 2024). The framework addresses RAI governance through four core functions: *Govern*, *Map*, *Measure*, and *Manage*. The *Govern* function addresses risk management. For example, *Govern* activities include developing accountability structures, setting up processes for stakeholder engagement, and improving diversity and culture. The *Map* function focuses on identifying the benefits and risks of AI systems based on their intended use. Example practices include checking assumptions and understanding the benefits and limitations of AI use for a specific use context. The *Measure* function applies relevant metrics to analyze, assess, benchmark, and monitor AI risks and impacts. The *Manage* function implements practices to reduce risk and increase positive impacts. Practices include creating protocols that prioritize, respond to, and manage activities in the Map and Measure functions, such as anticipating and managing downstream harms to end users (e.g, test takers).

---

[2] NIST AI RFM (2023) is a US framework. We acknowledge that RAI frameworks are available in other countries. However, we chose this framework for purposes of illustration.

The NIST AI RMF's *trustworthiness characteristics* (i.e., ethical pillars) draw from the International Organization for Standardization (ISO) and align with AI ethics literature (such as Fjeld et al., 2020; Jobin et al., 2019). They emphasize that AI systems should be: *Valid and Reliable; Safe, Secure and Resilient; Accountable and Transparent; Explainable and Interpretable;* and*, Privacy-Enhanced, Fair – with Harmful Bias Managed*. The trustworthiness characteristics interact across the four governance functions. For example, AI-powered assessments that are secure (*Govern*) also need to be *fair*; systems that are accurate (*Measure*) also need to be *interpretable*.

Chapelle and Voss (2021) assert that as we see advances in technology for assessment, assessment stakeholders should revisit the language constructs measured and implications for test score interpretation. This position suggests that we need to think about how to identify and measure responses to innovative task types (e.g., Attali et al., 2023; Goodwin et al., 2024; Runge et al., 2024). We need to consider the role that RAI practices play in the design and measurement of innovative item types (Burstein et al., 2024).  Taking this a step further, we need to address new security issues that have arisen with generative AI, such as methods for capturing bad faith use of ChatGPT for writing responses (Khalil & Er, 2023; Niu et al., 2024). With this in mind, it is important for assessment stakeholders to be thinking about the design and implementation of RAI frameworks and practices.

## The Evolution of  Responsible AI for Assessment

The term "responsible AI" was not commonly used when AI was first introduced for high-stakes assessment. However, RAI practices have been applied since that time. Widespread availability of generative AI for assessment has increased focus on RAI. Consistent with this trend, as noted earlier, RAI policy guidelines have proliferated in the broader education and assessment communities. This section reviews RAI theory and practice for assessment over time. The discussion highlights how earlier AI-powered assessment attended to responsibility by addressing foundational assessment validation theory inferences as conceptualized in Kane (1992, 2013) and Chapelle et al. (2008). We reference the six inferences introduced earlier: *domain definition, evaluation, generalization, explanation, extrapolation,* and *utilization.*[3]

The earliest, most well-researched, and mature use of AI in assessment is AWE. Project Essay Grade (PEG[4]; Page, 1966) was the first AWE system designed for scoring student writing. Page's motivation was to reduce the time that teachers spent on grading writing assignments. This early method relied on a word count measure (i.e., the fourth root of essay word count). With regard to validity argument inferences, Page's work[5] addressed *domain definition* by identifying *essay writing* as the relevant construct and *evaluation* by measuring agreement between human judge and system scores. Technology and computing infrastructure were limited at the time PEG was introduced which constrained wider adoption and a rigorous research agenda.

Through expanded research efforts in the computational linguistics and educational measurement communities, AWE development and use grew, including the use of natural language processing[6] (NLP). NLP methods were used to automatically identify linguistic characteristics in text-based writing samples beyond length-based features (Attali & Burstein, 2006; Burstein et al., 1998; Foltz et al., 1999; Shermis & Burstein, 2003). Coupled with statistical modeling and machine learning approaches, AWE systems showed great promise for large-scale, high-stakes assessment – a context in which responsible AI practices are essential. The remainder of this discussion outlines how theory and practice promoting responsibility evolved following the AWE use for high-stakes assessment.

---

[3] Note that some of the examples pre-date argument-based validity frameworks but address the inference concepts.
[4] The PEG system was the foundational system for [Measurement Incorporated](Measurement Incorporated).
[5] Validity research about the commercial version of PEG can be found here: https://www.measurementinc.com/research.
[6] See [Jurafsky and Martin (2023)](Jurafsky and Martin (2023))

The first NLP-based AES system introduced to high-stakes assessment was Educational Testing Service's (ETS) e-rater[®] scoring engine (Burstein et al., 1998). Three key RAI assessment practices emerged in Burstein et al., (1998) and Attali and Burstein (2006). These practices addressed the *domain definition*, *evaluation*, and *explanation* inferences. Burstein et al. discuss the writing skills (*domain definition*) that the system aimed to measure, and the relationships between the writing construct elements (e.g., vocabulary use, argument development) and the AES system features (*explanation*). They also present evaluations between the AES writing score predictions and human rater scores (*evaluation*). Attali and Burstein conducted factor analysis to provide empirical explanation on how the AES features align with the construct (*explanation*). Powers et al. (2002) conducted a study to assess the extent to which the earlier version of e-rater (Burstein et al., 1998) could be "stumped" by opening a challenge where users could try to trick the system. The idea was to identify threats to validity – specifically, users could submit writing to e-rater with the intention of using techniques to "trick" the system into assigning higher or lower scores. This research demonstrated how construct-irrelevant variables (e.g., lengthy but repetitive essays) impacted the system scores and could threaten the explanation inference (i.e., violating construct relevance). In the context of explanation, Cahill et al. (2018) and Higgins et al. (2006) introduced novel approaches to identifying and flagging "bad faith" essays on writing assessments to mitigate potential threats to test validity. In cases of bad faith writing, test takers might engage in strategies such as repeating the prompt text multiple times to "write" a long essay, or trying to fool the system by introducing rare but irrelevant vocabulary. These "bad faith" activities violate the explanation inference in that the writing does not represent the test-taker's ability to complete the assigned task.

Williamson et al.'s (2012) theoretical paper leveraged Chapelle et al. (2008) to outline what we would now refer to as RAI practices. Focussing on automated scoring used for written and spoken constructed responses, Williamson et al.'s recommended practices support a validity argument, while also mitigating threats to test validity. The paper aimed to provide a validation framework with systematic guidelines for the responsible use of automated scoring for assessment. It discusses relevant activities associated with Chapelle's inferences to build a validity argument for an assessment. Five inferences are explicitly addressed: *explanation* (e.g., construct evaluation and rubric criteria); *evaluation* (e.g., human-system agreement); *extrapolation* (e.g., empirical performance between automated scores and independent measures); *generalization* (e.g., reliability on alternate tasks), and *utilization* (e.g., disclosure about strengths and limitations of automated scoring to inform score users).

As the number of commercial AES systems grew, there was broad interest from educational stakeholders in an independent, systematic comparison to ensure validity – aligned with Chapelle et al.'s evaluation inference. Shermis and Hamner (2013) report on the first such *competitive evaluation* ("bake-off") of nine systems (i.e., eight commercial and one open source system) including ETS's e-rater, Pearson's Intelligent Essay Assessor, and Measurement Inc.'s PEG[7]. The competition used eight different essay data sets from U.S. K-12 writing assessments. Data sets varied with regard to genre, including source-based, persuasive, expository, and narrative writing samples. Data set sizes ranged from approximately 1,000 to 3,000 responses which were used for model training and evaluation. The study focussed on the *evaluation* inference; to compare system performance, the evaluation was conducted by examining human-system agreement on various metrics, including exact and adjacent percent agreement and Kappa. While the study had limitations, some of which included no consideration of construct validity (*explanation*) and generalizability (*generalization*; Shermis & Hamner, 2013), in the spirit of RAI, it was the first to conduct and disseminate an independent evaluation of competing commercial AI-powered automated writing evaluation systems.

---

[7] Refer to Shermis and Hamner (2013) for more details about the different writing evaluation systems.

Chapters in Shermis and Burstein (2003, 2013), Yan et al. (2020), and Shermis and Wilson (2024) primarily address topics related to *explanation*, *evaluation*, and *generalization* of automated scoring systems over 20 years. These collective volumes illustrate the assessment research community's dedication to and evolution of responsible AI theory and practice, with a focus on automated writing and speech scoring.

Johnson et al. (2022) offer a detailed discussion of bias in automated scoring of written and spoken responses. This is the first paper, to our knowledge, that drills down to discuss seven categories of bias that need attention with regard to validity research for automated scoring. The categories include *Historical bias* (i.e, historical data not representative of current populations)*, Representation bias* (i.e., data underrepresentation)*, Measurement bias* (e.g., omitted or proxy variables)*, Aggregation bias* (i.e., aggregated data applied to subgroups), *Learning/Algorithmic bias* (i.e., algorithms that impact subgroups), *Evaluation bias* (e.g., choice of groups to evaluate fairness), and *Deployment bias* (i.e., algorithmic outputs used in a manner inconsistent with algorithms used to produce them, such as rounding or truncation). Johnson et al. highlight the critical need to attend to the details in algorithmic bias research to ensure fairness for automated scoring for writing and speaking assessment (*evaluation*).

AERA, APA, and NCME (2014) published professional assessment standards that addressed best practices for automated scoring of written and spoken constructed responses. These standards explicitly address assessment validation inferences associated with automated scoring: for example, addressing score bias (*evaluation*), linking scores to the construct (*explanation*), ensuring that score criteria were embedded in the machine scorers (*domain definition/explanation*), and building in quality control methods to evaluate machine scorers (*evaluation*).

More recent assessment guidelines and policy address responsible AI more directly and offer sufficient detail to cover the six inferences. The International Test Commission and Association of Test Publishers (ITC-ATP; 2022) developed guidelines for digital assessments. The guidelines comprehensively address aspects of test design, measurement, and security, with details associated with data governance that exceed discussions in the Standards. ATP (2024) published policy guidelines that explicitly address responsible AI for assessment. The guidelines address *organizational considerations*, such as governance, risk assessment, and change management protocols; *general AI principles* that speak to industry-agnostic practices, such as promotion of human values, human involvement, and data privacy; and *AI topics specific to assessment*, such as accessibility, test validity and reliability, and test-taker interactions with AI.

Burstein (2023) introduced the Duolingo English Test (DET) RAI standards for a language assessment. Consistent with the NIST AI RMF (2023) framework (discussed earlier), the DET RAI standards for language assessment align ethical principles with governance functions. For instance, conceptually aligned with the NIST AI RMF trustworthiness characteristic, "Valid and Reliable", Burstein's Validity and Reliability standard goal aims to evaluate construct relevance and accuracy (validity) and test score consistency (reliability).

Activities to uphold this standard include: aligning test item types to the test domain (*Map*); evaluating the test's scoring approaches for accuracy and fairness (*Measure*); and, identifying methods that produce valid and reliable test scores (*Manage*). In another example, Burstein's (2023) Accountability and Transparency standard aims to build trust with stakeholders. Activities include documenting how test takers interact with AI (*Map*), and documenting the required qualification for human experts involved in the development and deployment of AI (*Govern*). The six inferences are explicitly addressed in the Validity and Reliability standard. However, the additional standards also address inferences. For example, the Fairness standard directly addresses the bias (evaluation).

This section has illustrated how RAI has been attended to over time, and how it interacts with assessment validation inferences. However, there is currently no consensus RAI framework for assessment. In the next section, we demonstrate the complementary nature of a domain-agnostic RAI framework and RAI assessment practices.

**Integrating Classical Validation Theory and Responsible AI**

Our discussion has focussed, thus far, on classical validation theory and domain-agnostic RAI principles and practice. As AI is now prominently used for high-stakes test development, measurement, and security, this situation calls for a broader, unified approach that integrates assessment-specific and domain-agnostic RAI into validity theory. From an assessment perspective: *How do we integrate responsible AI principles and practice, and assessment to ensure test validity and attend to social responsibility?* To illustrate, we present the Duolingo English Test (DET) language assessment ecosystem as a model (Burstein et al., 2022). The remainder of this section provides an overview of the ecosystem, illustrating the integration of the *assessment-specific* and *domain-agnostic* factors that contribute to the test validity argument.

*The Duolingo English Test Assessment Ecosystem*

The DET's assessment ecosystem is composed of a set of integrated assessment frameworks for test design, measurement, and security. Test-taker experience (TTX) factors, RAI standards, and the "digital" chain of inferences are ecosystem components that interact with the frameworks. Figure 1 illustrates how the frameworks for test design, test measurement (*computational psychometrics*; von Davier, A., 2017, and the *expanded evidence-centered design*; Arieli-Attali et al., 2019), and test security work together. For example, as part of test design, we need to make sure that test security is considered. To mitigate cheating behaviors, for instance, we cannot design test items that allow test takers to use internet search, since this would involve working outside of the test environment, compromising test security. In another example, we want to make sure that test items support construct-relevant evidence collection to ensure that the test provides a score that is relevant to test-takers' skills and abilities.

With regard to TTX, for instance, test access is essential. It is important to ensure that the test is broadly accessible to all by maintaining a lower, more affordable test price. Specific to RAI, the DET has developed RAI Standards (Burstein, 2023). Practices associated with the RAI Standards are applied in test design, measurement, and security in an ethical way so as not to cause harm and ultimately support the validity argument – that is, the "digital" chain of inferences (DCI). Adapted from earlier validity argument frameworks (Chapelle et al., 2008; Kane, 1992, 2011), the ecosystem's DCI explicitly addresses the technology used across the ecosystem frameworks.

RAI practices can be *assessment-specific* and *domain-agnostic*. This characteristic is key to a unified assessment ecosystem that seamlessly integrates RAI principles and classical assessment validation. This integration is essential to mitigate direct and indirect harms related to the assessment. We consider direct harms to be those that are assessment-specific and impact the validity of the assessment: direct harms compromise validation inferences (e.g., introduce construct-irrelevant variance). By contrast, indirect harms are domain-agnostic, meaning that they do not directly affect test validity. However, they may contribute to broader social harms and have social responsibility implications for an assessment (e.g., environmental impact). This section discusses the role of assessment-specific and domain-agnostic RAI principles in a unified assessment ecosystem.

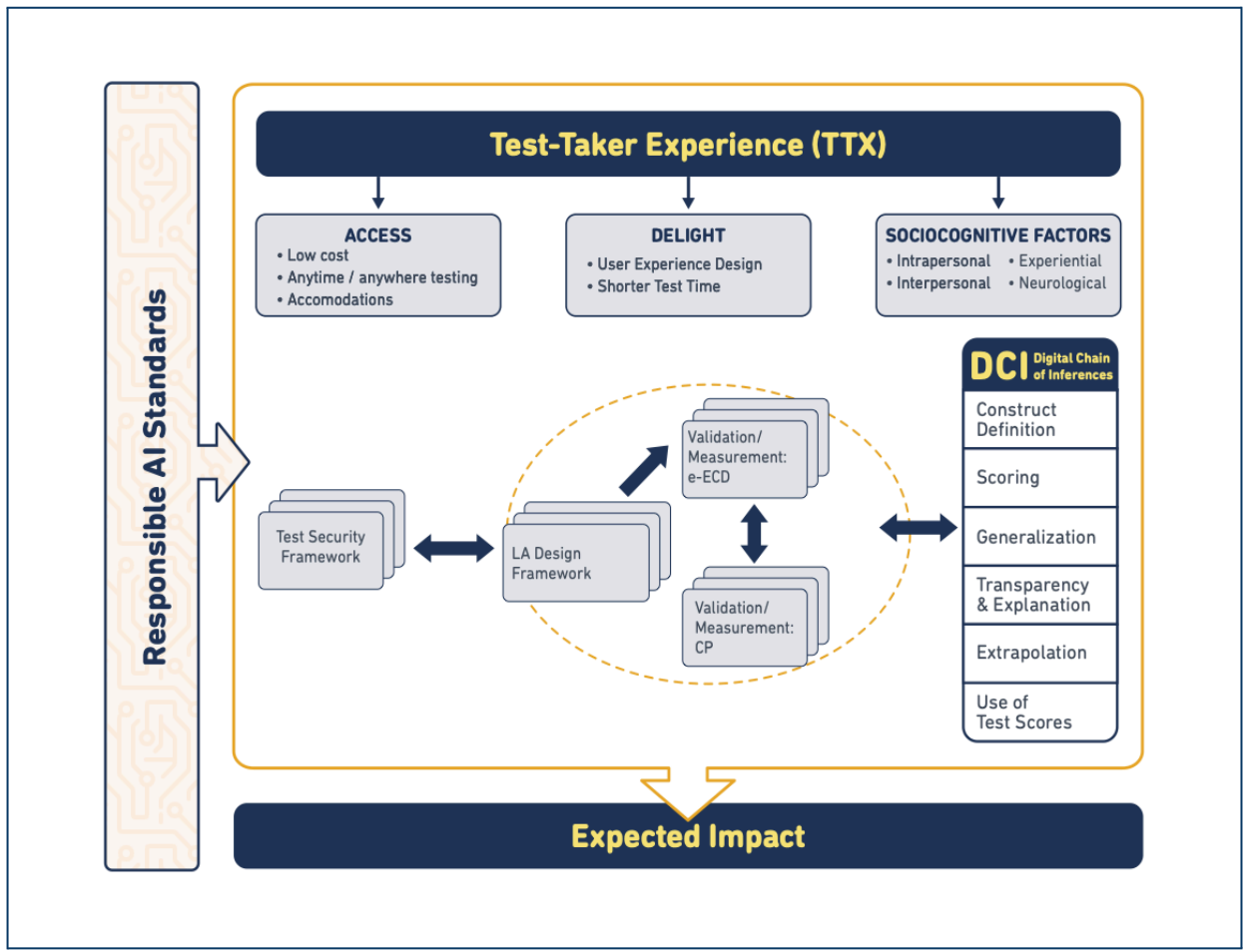

**Figure 1.** The DET assessment ecosystem (Burstein et al., 2022); e-ECD refers to the Expanded Evidence-Centered Design , and CP refers to Computational Psychometrics.

*Assessment-specific RAI practices*

RAI elements that are directly related to assessment are incorporated into the ecosystems frameworks in order to ensure that there is evidence for the various inferences in the validity argument that those frameworks support. Burstein et al. (2024) illustrate how the four RAI standards from Burstein (2023) are applied to develop a reading task and an automated writing scoring system, to ensure test content is appropriate and free from bias, and to support the security of test scores.

Connecting responsible AI standards to inferences in validity theory is relatively straightforward. A line can be drawn from the first two goals of the Validity and Reliability standard in Burstein (2023) to the *domain description* inference in a validity argument. For example, the first two Validity and Reliability standards include goals in which subject-matter experts outline the target domain of a task and the construct it is intended to measure. This step essentially creates a version of test specifications (Davidson & Lynch, 2002) that can be restructured into system prompts for the LLM to produce stimuli for reading and listening tasks. The large context windows of LLMs allow for large system prompts, giving test developers the opportunity to include a wide range of construct-relevant instructions in the system prompt. For example, specifications (or system prompts) for reading stimuli can provide information regarding text type, text length, sophistication of vocabulary, content alignment to a framework such as CEFR (Council of Europe, 2020), and topic/subject. In this approach, the system prompt is treated as the item specification that is given to the LLM (similar to an item writer) to generate content that is relevant to the target domain, which bolsters evidence for the domain description inference.

Similar to human-written content, generated content needs to be reviewed by humans to ensure that it meets requirements to ensure high-quality measurement of ability and to remove objectionable content (Burstein et al., 2024). This process aligns with steps in Burstein's (2023) fairness standard, which outlines steps to mitigate bias that can be found in large language models. These steps include human review processes as well as quantitative analyses to investigate for differential item functioning (Belzak et al., 2023).

As presented in Williamson (2012), Burstein (2023), and Burstein et al. (2024) responsible AI for automated scoring systems creates evidence that supports at least five inferences in a validity argument. Burstein et al. (2024) demonstrate how the development of construct-relevant rubrics used by humans to create "gold-standard" datasets for training automated scoring systems creates strong ties to the construct, which along with task design supports the *explanation* inference. Additionally, agreement between automated scoring systems and the gold standard generates evidence for the *scoring* inference in a validity argument.

Malicious behavior on the part of test takers can pose a threat to the validity of test scores. Burstein's Privacy and Security standard outlines steps to ensure that human-in-the-loop AI proctoring protocols can identify known cheating behaviors fairly and reliably. Niu et al. (2024) demonstrate how LLMs can be trained to detect when test takers copy-type outputs from LLMs as responses to test items with low false positive rates. However, despite the low false positive rates, the authors strongly recommend human oversight in using these models. In current validity frameworks, this generates evidence for the *utilization* inference as it demonstrates that test scores are attributable to the test taker and their proficiency.

The content above reframes the evidence that is generated from assessment-specific RAI processes. It provides examples of how RAI processes can support the *domain description*, *scoring*, *explanation*, and *utilization* inferences in a validity argument. The same logic could be extended to other inferences to demonstrate how, for instance, RAI practices could support feature-based item parameters or the development of a computer-adaptive testing algorithm – both of which contribute to evidence, for example, reliable test scores (i.e., *generalization* inference). Additionally, the examples provided above demonstrate how RAI practices can mitigate direct harms to the validity of test score interpretation and use. Further, we argue that it is important for users of AI in the development, administration, and scoring of language tests to consider indirect harms of AI, which can be mitigated through domain-agnostic RAI practices. Indirect harms may not have effects on the interpretation and uses of test scores, but they are nonetheless important as both RAI practices and the use of language assessment scores for decision-making are part of a broader social landscape.

*Domain-agnostic RAI practices*

Social responsibility includes factors such as *environmental impact* and *labor* concerns related to the use of AI and, specifically, generative AI.  These factors do not directly impact test score validity. However, practices used to create the AI models used for test development and scoring may have social impact, and therefore indirectly affect test scores.

The DET's use of AI for security allows remote, at-home, 24/7 testing versus testing-center assessments. Remote testing provides greater access to a more diverse set of test takers (Thomas & Belzak, 2024). With regard to *environmental impact*, the remote testing scenario does not require travel, thus potentially reducing carbon emissions associated with test-taker travel (e.g., car, train, airplane) to a test center. Resource-heavy computing resources required for generative AI models may have environmental impact (NIST AI RFM, 2023) carbon emissions which, in turn, contribute negatively to climate issues include, such as warmer temperatures contributing to disruptive weather conditions (United Nations, n.d.), and poor air quality that impacts human health risks (U.S. Environmental Protection Agency, 2024). Assessment organizations that use generative AI for test development, such as for automated item generation, should include RAI practices to monitor their computing resources associated with generative AI model use.

Especially with the widespread use of AI, *labor* has been a central discussion. This is not novel and thus has been discussed since early automation (Autor et al, 2024). Concerns have arisen, for instance, related to job loss (i.e., jobs replaced by AI) and the need for upskilling as many job roles may require AI skills (Green, 2024). With regard to assessment, early uses of AI (such as automated essay scoring) led to a reduction in the number of human raters (e.g., Burstein et al., 1998). More recent uses of AI rely on increased human oversight, including humans refining AI systems with examples to align outputs more closely with human expectations (Wang, 2019). In addition, recent education and assessment policy discusses human-in-the-loop AI in the context of human oversight at critical decision points (ATP, 2024; US Department of Education, 2023). The assessment community should remain cognizant of the critical need for human involvement (such as for fairness and bias review). As well, the community should be thinking about the new careers that may arise. To that end, AI literacy training also should be in place for assessment developers to learn new AI skills, such as prompt engineering, that can contribute to the item development process. The assessment community should be looking to the future of assessment to understand what AI skills will allow people to continue to contribute productively to assessment.

**Conclusion & Future Directions**

Expanding on Chapelle et al.'s (2008) foundational work, this paper proposes a unified approach that integrates classical test validation theory, and *assessment-specific* and *domain-agnostic* RAI principles and practice. The paper examines the relationship between classical assessment validation theory and RAI principles and practice. It further asserts that a broader view of RAI for assessment better ensures the ethical and valid use of AI in high-stakes assessments through the entire assessment ecosystem for test design, measurement (scoring), and secure test administration. As well, it encourages human oversight. Specifically, it advises stakeholders who use AI to design and evaluate language assessments, such as test developers and researchers, to think critically about how to implement responsible AI (von Davier, A. & Burstein, in press). This ensures that evidence collected from an AI-powered assessment supports one or more inferences in a validity argument. In addition, it highlights the importance of domain-agnostic RAI factors with regard to social responsibility such as the impact of AI model use on the environment and labor issues.

The concepts of validity in assessment and RAI share a common goal of leveraging an evidence-based design for the purpose of building trustworthy products and systems. AI models continue to improve in their current capabilities and AI-powered capabilities are developed at a rapid pace; this facilitates the integration of these capabilities (e.g., image generation or detection of deep fakes) into the language assessment ecosystem. Continuous innovation requires that users of the AI for assessment regularly reevaluate their assessment-specific and domain-agnostic RAI principles and practices, ensuring that RAI ideals are translated in practice.

As AI adoption for assessment becomes increasingly more prevalent, it is advisable for test developers and assessment researchers to evaluate the *maturity* of their RAI practices (Dotan et al., 2024). Specifically, there should be an emphasis on evaluating the extent to which RAI practices are developed, documented, and implemented. Attention to implementation is key, ensuring that RAI practices are systematically evaluated and applied to AI-powered assessments.

## Acknowledgements

We would like to acknowledge Ravit Dotan for her guidance about the alignment between the NIST AI RMF and the DET RAI Standards. Many thanks to our Duolingo colleagues, Ben Naismith, Yena Park and Alina von Davier for their reviews.